# Planning future charging infrastructure for private EVs: A city-scale assessment of demand and capacity

Hong Yuan [1], Minda Ma [2*, 3], Nan Zhou [4], Yanqiao Deng [1], Junhong Liu [2], Shufan Zhang [1], Zhili Ma [1]

1. School of Management Science and Real Estate, Chongqing University, Chongqing, 400045, P. R. China
2. School of Architecture and Urban Planning, Chongqing University, Chongqing, 400045, P. R. China
3. Building Technology and Urban Systems Division, Energy Technologies Area, Lawrence Berkeley National Laboratory, Berkeley, CA 94720, United States
4. Energy and Resources Group, University of California, Berkeley, CA 94720, United States

- Corresponding author: Prof. Dr. Minda Ma, Email: maminda@lbl.gov
  Homepage: https://globe2060.org/MindaMa/

## Graphical abstract

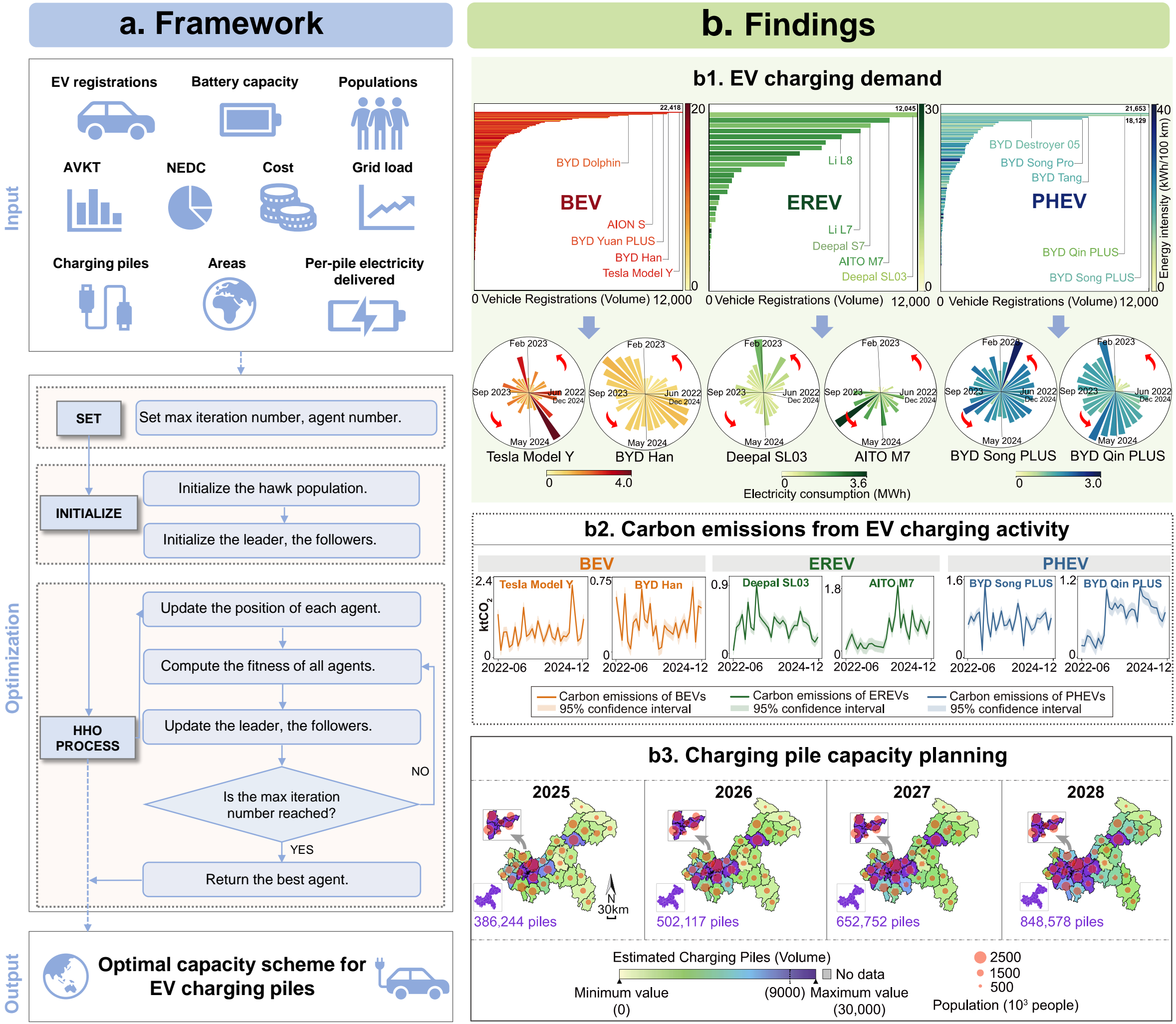

**Graphical Abstract:** A city-scale capacity planning model for EV charging infrastructure and a case study of Chongqing. The graph presents (a) the operational framework of the planning model; (b1) the distributions of energy intensity and model-specific electricity consumption for BEVs, EREVs, and PHEVs; (b2) monthly variations in operational carbon emissions from top-selling EV models; and (b3) the spatial allocation of charging pile capacity across Chongqing's districts and counties.

## Highlights

- Proposed a globally adaptable, city-scale capacity planning model for EV charging infrastructure.
- EV electricity consumption in Chongqing reached 57.5 GWh in Dec 2024, resulting in 30 kt $CO_2$ emissions.
- Installed 181,622 charging piles during 2022–2024, with spatial distribution aligned to population density.
- By 2030, EV electricity demand is projected to reach 1940 GWh, with charging piles exceeding 1.3 million.
- Developed a versatile tool for stakeholders to plan sustainable, cost-effective, low-carbon EV infrastructure.

## Abstract

Capacity planning for electric vehicle (EV) charging infrastructure has emerged as a critical challenge in developing low-carbon urban energy systems. This study proposes the first demand-driven, multi-objective planning model for optimizing city-scale capacity allocation of EV charging infrastructure. The model employs a bottom-up approach to estimate charging demand differentiated by vehicle type—battery electric vehicles (BEVs), extended-range electric vehicles (EREVs), and plug-in hybrid electric vehicles (PHEVs). Chongqing, a rapidly expanding EV industry cluster in China with a strong industrial base, supportive policies, and diverse urban morphologies, is selected as the case study. The results show that (1) monthly EV electricity consumption in Chongqing rose from 18.9 gigawatt-hours (GWh) in June 2022 to 57.5 GWh in December 2024, with associated carbon emissions increasing from 9.9 kilotons of carbon dioxide (kt$CO_2$) to 30 kt$CO_2$; (2) 181,622 additional charging piles were installed between 2022 and 2024, with the fastest growth observed in Yubei, reflecting a demand-responsive strategy that prioritizes areas with higher population density, higher income levels, and adequate land availability for pile deployment, rather than broad geographic coverage; and (3) between 2025 and 2030, EV electricity demand is projected to reach 1940 GWh, with the number of charging piles exceeding 1.4 million, and charging demand from EREVs and PHEVs expected to overtake BEVs later in the period. While Chongqing serves as the pilot area, the proposed planning platform is adaptable for application in cities worldwide, enabling cross-regional comparisons under diverse socio-economic, geographic, and policy conditions. Overall, this work offers policymakers a versatile tool to support sustainable, cost-effective EV infrastructure deployment aligned with low-carbon electrification targets in the transportation sector.

## Keywords

Electric vehicles; Charging infrastructure; Electricity demand; Capacity planning; Multi-objective optimization

## Abbreviations and symbols

Considering the space limitation, please refer to Appendix C for this part.

## 1. Introduction

### *1.1. Background*

The global shift toward electric mobility is accelerating, driven by climate targets and growing concerns over air pollution and energy security [1, 2]. As efforts to decarbonize the transport sector intensify, cities (particularly in emerging economies) face increasing pressure to expand charging infrastructure in step with increasing electric vehicle (EV) adoption [3]. The International Energy Agency projects that EVs will constitute more than 55% of vehicle sales worldwide by 2035 under net-zero scenarios [4], highlighting the urgency of scalable and cost-effective infrastructure strategies. As the world's largest EV market, China has installed more than 12 million charging piles by 2024 [5]; however, regional disparities remain due to uneven development and localized demand. City-scale planning is further constrained by the lack of efficient demand estimation tools [6, 7]. As a rapidly growing city experiencing swift EV adoption and evolving transportation dynamics, Chongqing exemplifies the planning challenges faced by cities in transition and serves as a critical testbed for EV infrastructure development.

### *1.2. Literature review*

As transportation electrification accelerates, accurately estimating EV charging demand becomes crucial for ensuring the reliability, efficiency, and equity of charging infrastructure. Most existing capacity planning models rely on simplified demand projections that overlook the heterogeneity among powertrain types [8]. Although some studies have analyzed battery electric vehicles (BEVs) separately or have aggregated all EVs into a single category [9], few have explicitly differentiated between powertrain systems in terms of energy consumption or mileage [10]. This implies that uniform assumptions can distort city-scale demand estimates and may hinder equitable infrastructure deployment. Therefore, a more data-informed and systematic capacity planning framework is urgently needed to better align charging infrastructure development with real mobility patterns and decarbonization goals.

Current approaches for estimating EV charging demand can be broadly categorized into data-driven models and simulation-based models, the latter encompassing both system-level and bottom-up frameworks. Data-driven models (such as deep learning [11], large language models [12], and time-series regressions [13]) leverage historical or real-time data to capture spatiotemporal charging patterns but often require extensive data inputs and face generalizability constraints. Alternatively, system-level models [14] simulate charging dynamics on the basis of traffic flows, grid interactions, or behavioral scenarios, providing system-wide insights but relying on predefined parameters and potentially overlooking local variations [15]. In contrast, bottom-up approaches estimate demand by aggregating individual vehicle-level charging behavior [16, 17], offering more accurate estimates across powertrain types and urban forms, particularly in cities where data limitations or demand heterogeneity pose challenges for top-down estimation.

Several studies have investigated charging infrastructure planning by developing models that optimize siting and sizing decisions under various spatial and technical constraints [18, 19]. Most adopt localized frameworks that target charging station deployment within communities [20], campuses [21], or sub-city districts [22]. Location–allocation models are commonly used to maximize service coverage or minimize user detours, considering land use [23], traffic density [24], and accessibility metrics [25]. In parallel, other studies have incorporated grid-side constraints, such as transformer capacity and voltage deviation, into charging station layout models to mitigate local load impacts [26, 27]. In addition to these spatial objectives, many studies have introduced cost-oriented goals, such as minimizing total investment and operational costs [28] or improving station utilization efficiency [29]. Furthermore, recent work has incorporated fairness and equity metrics to ensure that charging accessibility is evenly distributed across urban areas, preventing overconcentration of resources in high-demand zones [30]. A related line of research has addressed the placement of highway or intercity fast-charging stations using flow refueling location models that simulate vehicle routing and charging behavior [31, 32]. These methods have provided valuable insights at specific scales. However, most existing studies have primarily focused on micro-level design, station-level operations, or single-objective optimization [33, 34], offering limited capability to capture trade-offs among cost, grid stability, and spatial equity at the city scale—especially under long-term electrification and decarbonization goals.

To solve planning models, various optimization methods have been employed [35, 36]. Traditional techniques include mixed-integer linear programming [37] and dynamic programming [38], which are effective for small-scale or deterministic problems but face limitations in terms of complexity and uncertainty. Consequently, heuristic and metaheuristic algorithms (such as genetic algorithms [39], particle swarm optimization [40], and simulated annealing [41]) have gained popularity because of their flexibility and global search capabilities. More recent work has adopted bioinspired methods such as the Harris Hawks Optimizer (HHO) [42], which offer faster convergence and greater solution diversity in nonlinear, multi-objective scenarios. These developments reflect a shift toward more adaptive and scalable approaches for real-world EV infrastructure planning.

### *1.3. Motivation, contributions, and the organization*

On the basis of the above review, two major research gaps hinder effective planning for EV infrastructure under large-scale electrification. First, while prior studies have advanced EV charging demand estimation, most rely on aggregate-level projections that overlook powertrain differences and offer limited insight into demand at the individual vehicle level [43, 44]. Second, existing studies have focused primarily on charging stations or community-level deployment, and comprehensive planning at the city scale remains largely unexplored [45, 46]. Given the rapid pace of city electrification, city-wide planning must account for both central and peripheral districts, balancing cost, grid stability, and spatial equity [47]. Addressing these limitations is crucial for developing scalable and adaptable infrastructure strategies that align with long-term decarbonization goals. Three key research questions thus arise from these gaps and warrant further investigation:

- How can EV charging demand and emissions be estimated across powertrain types?
- How will citywide EV charging demand and charging infrastructure evolve over time?
- What is the optimal city-scale strategy for allocating future charging pile capacity?

To address the identified challenges, this study develops a demand-driven planning framework for allocating EV charging capacity at the city scale. First, a bottom-up approach is applied to estimate charging demand with seasonal-to-monthly temporal resolution, disaggregated by powertrain type—including BEVs, extended-range electric vehicles (EREVs), and plug-in hybrid electric vehicles (PHEVs). On the basis of these estimates, a citywide multi-objective optimization model is developed to determine the optimal charging infrastructure capacity for 2022–2030, balancing cost minimization with load leveling. In particular, the model is solved via the HHO algorithm, which enhances computational efficiency and adaptability under large-scale, real-world conditions. Chongqing is selected as a representative case study, given its status as a rapidly growing EV industry cluster[1] in China, supported by a strong industrial base, favorable policies, and diverse urban morphology—offering transferable insights for city-scale infrastructure planning in other urban contexts.

**The principal contribution of this study** lies in its development of a transferable, demand-driven planning framework for EV charging infrastructure that can be applied across diverse urban contexts globally. By integrating electricity consumption at the individual vehicle level, the bottom-up demand estimation approach enhances granularity and accuracy over conventional aggregate models. The optimization framework further incorporates both technical and equity considerations, enabling more inclusive infrastructure strategies that balance grid constraints and cost efficiency. This study addresses a critical gap in existing studies, which either focused narrowly on community-level charging piles or adopted coarse assumptions about aggregate demand, limiting their applicability to citywide infrastructure planning. While Chongqing provides a demonstrative case, the methodology is designed for broader applicability in cities facing rapid electrification and urban transformation. As such, this study offers a scalable tool for supporting long-term infrastructure deployment, scenario evaluation, and policy design in diverse urban contexts.

The remainder of this study is organized as follows: Section 2 details the construction of the modeling framework, including the bottom-up estimation of vehicle-type-specific charging demand and the development of the capacity planning model, and presents the data sources. Section 3 presents a historical charging demand analysis, associated carbon emissions, and an optimal capacity allocation scheme for EV charging piles. Section 4 provides a forward-looking capacity planning analysis for charging piles. Finally, Section 5 summarizes the key findings and outlines directions for future research.

---

[1] While Chongqing serves as a major EV industrial cluster, this study focuses on charging demand from private EVs. Industrial and public transport sectors are excluded as they typically rely on closed-loop, depot-based charging systems with distinct operational schedules and infrastructure needs.

## 2. Materials and methods

This study developed an integrated methodological framework to estimate EV charging demand, assess associated carbon emissions, and optimize city-scale infrastructure capacity planning. Section 2.1 establishes a bottom-up model for estimating charging demand and emissions across different EV powertrain types (including BEVs, EREVs, and PHEVs). Section 2.2 presents a capacity planning model for EV charging piles, which formulates a multi-objective optimization problem that balances cost and load fluctuations and is solved via the HHO algorithm. Section 2.3 describes the main data sources.

### *2.1. Bottom-up model for estimating EV charging demand and associated emissions*

Building on previous studies (e.g., [10, 16]) and accounting for seasonal variability in EV energy consumption—particularly the increased electricity use associated with cabin heating and cooling—the quarterly electricity consumption for each EV model is defined as follows:

$$\omega_j = Registrations_j \cdot \frac{Mileage_k \cdot Capacity_j}{\lambda_j \cdot NEDC_j} \tag{1}$$

$$\upsilon_j = Registrations_j \cdot \frac{Mileage_k \cdot Capacity_j}{\rho_j \cdot NEDC_j} \tag{2}$$

where $\omega_j$ [kilowatt-hour (kWh)] represents the total electricity consumption of model *j* (e.g., Tesla Model Y) in seasons without cabin heating or cooling; $\upsilon_j$ (kWh) represents the total electricity consumption of model *j* in seasons with cabin heating or cooling; $Mileage_k$ [kilometer (km)/year] signifies the average mileage in city *k* (e.g., Chongqing); $Capacity_j$ (kWh) denotes the battery capacity of model *j*; $NEDC_j$ (km) represents the New European Driving Cycle (NEDC) value of model *j* (A sensitivity analysis comparing results across different driving cycles is provided in Appendix D); $\lambda_j$ reflects the NEDC degradation coefficient of model *j* in seasons without cabin heating or cooling; $\rho_j$ reflects the NEDC degradation coefficient of model *j* in seasons with cabin heating or cooling; and $Registrations_j$ pertains to the registrations of model *j*. Therefore, the annual electricity consumption of model *j* can be expressed as follows:

$$\Omega_j = \omega_j + \upsilon_j \tag{3}$$

where $\Omega_j$ (kWh) represents the annual electricity consumption of model *j*. Hence, the annual electricity consumption of EV operations in city *k* is defined as follows:

$$E_k = \sum_{j=1}^{m} \Omega_j \tag{4}$$

where $E_k$ (kWh) denotes the annual electricity consumption of EV operations in city *k*. Finally, Eq. (4) can be expressed as follows:

$$E_k = \sum_{j=1}^{m} Registrations_j \cdot \frac{Mileage_k \cdot Capacity_j}{\lambda_j \cdot NEDC_j} + Registrations_j \cdot \frac{Mileage_k \cdot Capacity_j}{\rho_j \cdot NEDC_j} \tag{5}$$

Given the carbon emission factor of the power sector in city *k* and the electricity consumption of model *j*, the operational-phase carbon emissions of model *j* can be calculated as follows:

$$C_j = \eta \cdot \Omega_j \tag{6}$$

where $C_j$ [kilogram of carbon dioxide ($kgCO_2$)] denotes the total operational-phase carbon emissions of model *j*, and where $\eta$ ($kgCO_2$/kWh) is the emission factor in the power sector of city *k*. Accordingly, the total operational-phase carbon emissions of EVs in city *k* can be expressed as the sum of emissions across all vehicle models:

$$C_k = \sum_{j=1}^{m} C_j = \sum_{j=1}^{m} \eta \cdot \Omega_j \tag{7}$$

### *2.2. Capacity planning model for EV charging piles*

Considering the minimization of the total cost of deploying EV charging piles, the first objective function of the capacity planning model for piles is defined as follows:

$$min\ f_1 = min \sum_{i=1}^{N} \left( C_{cap,i} \cdot CRF + C_{om,i} \right) \cdot x_i \tag{8}$$

where $N$ is the total number of administrative districts in city *k*; $x_i$ (piles) denotes the number of charging piles allocated to district $i$; $C_{\text{cap},i}$ [Chinese Yuan (CNY, current price)/pile] represents the initial investment cost per charging pile in district $i$; $C_{\text{om},i}$ (CNY/pile/year) is the annual operation and maintenance cost per charging pile in district $i$; and CRF is the capital recovery factor, defined as:

$$CRF = \frac{h(1+h)^n}{(1+h)^n - 1} \tag{9}$$

where $h$ denotes the discount rate and $n$ (years) represents the project lifetime. Considering the minimization of grid load fluctuations on a quarterly basis, the second objective function of the capacity planning model for EV charging piles is formulated as follows:

$$min\ f_2 = min \frac{1}{4}\sum_{s=1}^{4}(Load_s - AvgLoad)^2 \tag{10}$$

where $Load_s$ (kW) denotes the total grid load in quarter $s$ and where $AvgLoad$ (kW) represents the average load over the entire year. On the basis of the above two objective functions, a multi-objective optimization problem was formulated to minimize the total life-cycle cost of the charging infrastructure and the fluctuation of the grid load:

$$minF = min\ (f_1 + f_2) \tag{11}$$

Considering that the cost ($f_1$) and load fluctuation ($f_2$) have different units and magnitudes, both objectives were first normalized to dimensionless ranges to ensure comparability:

$$f_i^{'} = \frac{f_i - f_i^{min}}{f_i^{max} - f_i^{min}}, \quad i = 1,2 \tag{12}$$

where $f_i^{min}$ and $f_i^{max}$ denote the minimum and maximum values of each objective. Subsequently, the normalized objectives were aggregated into a single equivalent function through a weighted sum approach, which is a common method for solving multi-objective optimization problems:

$$F = w_1 f_1^{'} + w_2 f_2^{'}, \quad w_1 + w_2 = 1 \tag{13}$$

where $w_1$ and $w_2$ denote the relative importance assigned to the total cost and grid load fluctuation, respectively. Instead of assigning subjective weights, this study adopted an entropy weighting method to determine the weights objectively.

Additionally, by comprehensively considering both technical and nontechnical factors, this study proposed the following six constraints for EV charging pile capacity planning:

***a.*** Total number of charging piles constraint:

$$\sum_{i=1}^{N} x_i = X_{total} \tag{14}$$

where $X_{\text{total}}$ denotes the total number of EV charging piles planned for city *k*.

***b.*** Quarterly charging demand satisfaction constraint:

$$\sum_{i=1}^{N} x_i \cdot \lambda_{i,s} \geq Q_s, \quad \forall s \in 1,2,3,4 \tag{15}$$

where $\lambda_{i,s}$ (kWh/pile/quarter) denotes the average energy delivered per pile in district $i$ during quarter $s$ and where $Q_s$ (kWh/quarter) is the total city-wide charging demand in quarter $s$.

***c.*** Quarterly grid load upper bound constraint:

$$\sum_{i=1}^{N} x_i \cdot p_{i,s} \leq \text{GridMargin}_s, \quad \forall s \in 1,2,3,4 \tag{16}$$

where $GridMargin_s$ (kW) represents the allowable additional grid load capacity for charging in quarter $s$ and where $p_{i,s}$ (kW/pile) denotes the average charging power per pile in district $i$ during quarter $s$.

***d.*** Equity constraints on regional allocation:

$$\left| \frac{x_i}{X_{\text{total}}} - \frac{D_i}{D_{\text{total}}} \right| \leq \delta, \quad \forall i \tag{17}$$

where $D_i$ ($10^3$ people) and $D_{total}$ ($10^3$ people) are the population of district $i$ and the total city population, respectively, and where $\delta$ is the allocation deviation threshold.

***e.*** Minimum charging pile density constraint:

$$\frac{x_i}{\text{Area}_i} \geq \alpha, \quad \forall i \tag{18}$$

where $\text{Area}_i$ [square kilometer ($km^2$)] is the area size of district $i$, and $\alpha$ (pile/$km^2$) is the minimum acceptable pile density.

***f.*** Priority guarantee for key area constraints:

$$x_t \geq \beta \cdot \frac{X_{total}}{N}, \quad \forall t \in \text{key areas (e.g., central urban districts)} \tag{19}$$

where $\beta$ is the allocation coefficient for prioritized areas. The default values, data sources, and reasonable ranges for $\delta$, $\alpha$, and $\beta$ are provided in Appendix E. Given the proposed capacity planning model, this study employed the HHO algorithm [48] to identify optimal solutions. The HHO algorithm is a bioinspired metaheuristic that mimics the cooperative hunting behavior and chasing strategies of Harris hawks when capturing prey (The algorithmic framework is illustrated in Appendix F). It adaptively balances exploration and exploitation, demonstrating fast convergence, strong global search capability, and robustness in solving nonlinear, high-dimensional problems. Compared with traditional algorithms such as particle swarm optimization, HHO achieves higher accuracy and efficiency, making it particularly suitable for complex EV charging infrastructure planning.

The algorithm operates in two main phases—exploration and exploitation—depending on the escaping energy $|E|$ of the prey. In the exploration phase, hawks selected a habitat on the basis of a random parameter $q$ to monitor and search for prey. As the prey escape energy decreases, the algorithm transitions to the exploitation phase, during which hawks initiated an attack on the basis of prior observations. The prey response was modeled via a randomly generated escape probability $r$, where $r < 0.5$ indicates successful escape and where $r \geq 0.5$ represents a failed attempt. Based on the values of $r$ and $|E|$, four distinct position update strategies were employed in the exploitation phase to simulate diverse pursuit behaviors and enhance convergence performance. Moreover, the modeling procedures and parameter settings for the future projections of charging demand and capacity planning are provided in Appendices G & H.

### *2.3. Data sources*

Chongqing, a rapidly expanding EV industry cluster in China with a strong industrial base, supportive policies, and diverse urban morphologies, was selected as the case area. This study examined all EV models registered in Chongqing over a three-year period (2022–2024), including BEVs, PHEVs, and EREVs. The objective was to evaluate electricity consumption and the associated carbon emissions during the operational phase and to derive the optimal capacity planning scheme for the EV charging infrastructure. A multisource data strategy was adopted to ensure accuracy and representativeness. EV registration volumes were obtained from the Home of Car Owners (https://www.16888.com/), a platform that provides detailed and regularly updated vehicle registration statistics in China. Model-specific technical data—including battery capacity (kWh) and NEDC (km)—were collected from Autohome

(https://www.autohome.com.cn/), which compiled specifications from manufacturers and official sources. Estimates of average annual mileage specific to EV users in Chongqing were drawn from an empirical study conducted by Ou et al. [49], which offered localized behavioral insights. Indirect carbon emissions from electricity consumption were calculated via regional grid emission factors published by Zhuo et al. [50]. The input parameters required for the capacity planning model in Chongqing (such as charging pile stocks, charging pile costs, charging power per pile, population, and land area) were obtained from official reports released by the Chongqing Municipal Government (https://cq.gov.cn/). The total grid load and the allowable additional grid load for EV charging were obtained from the State Grid Chongqing Electric Power Company (http://www.cq.sgcc.com.cn/). By integrating these diverse sources, this study offers a robust and context-specific assessment of EV charging demand and emissions in Chongqing, along with a data-driven foundation for optimal charging infrastructure capacity planning.

## 3. Results

Building on the methodological framework outlined in Section 2, this section presents three key empirical results on the basis of a case study of Chongqing. First, it evaluated the historical operational electricity consumption of EV models in Chongqing from 2022 to 2024, disaggregated by powertrain type. Second, it quantified the carbon emissions associated with EV charging over the same period. Third, the optimal historical allocation scheme for EV charging pile capacity across districts was presented.

### *3.1. Historical assessment of operational electricity consumption for EV models*

This study employed a bottom-up approach to assess the historical operation electricity consumption of EV models (BEVs, EREVs, and PHEVs) in Chongqing from June 2022 to December 2024. This study characterized the temporal evolution of EV charging demand at the city scale. This historical evaluation provided the foundation for subsequent emissions accounting and infrastructure planning. Fig. 1 presents a detailed analysis of the registration trends of different powertrain EV types, along with corresponding energy intensities, providing insights into market dynamics, technological performance, and consumer preferences. Fig. 1 a–c show the registration and energy intensity of BEVs, EREVs, and PHEVs, respectively, each revealing distinct patterns. In Fig. 1 a, the best-selling BEV models, such as Tesla Model 3 and BYD Han, generally demonstrated lower energy intensity. This trend reveals that BEV consumers tend to favor vehicles that balance market popularity with energy efficiency while also reflecting the effectiveness of manufacturers in promoting models with optimized energy management. The broad spectrum of registrations—from niche models with low uptake to mass-market models with high adoption—highlights the diversity of the BEV segment, where energy efficiency serves as a critical point of differentiation.

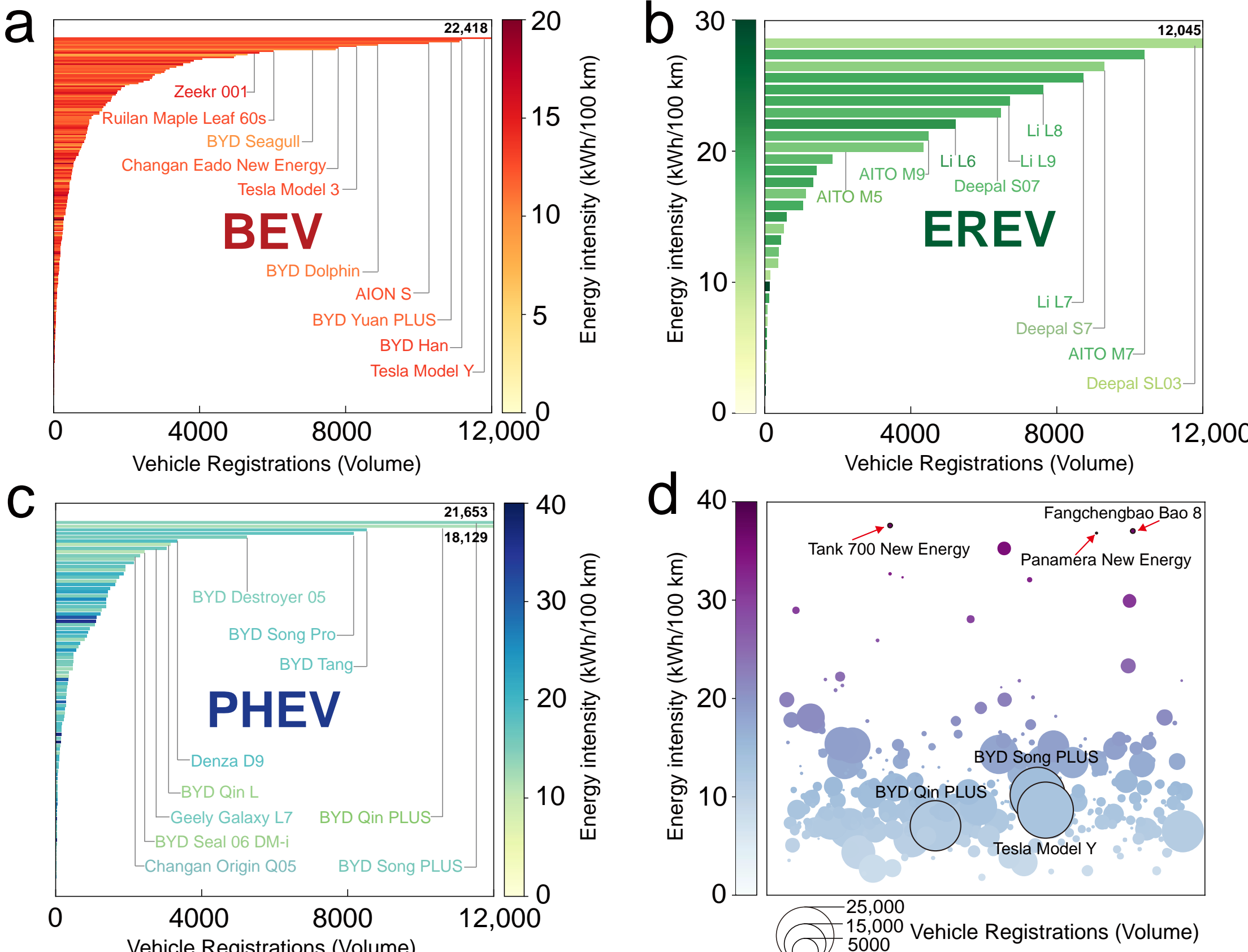

**Fig. 1.** Registration trends of different EV types from June 2022 to December 2024, along with corresponding energy intensities, for the case study area of Chongqing: (a–c) distributions of vehicle registrations and energy intensities for BEVs, EREVs, and PHEVs; (d) aggregated analysis of energy intensity and registrations across all EV categories. [2]

For the EREVs, Fig. 1 b reveals a similar yet distinct pattern. The top-selling models, such as Li L8 and AITO M5, exhibited moderate energy intensity, clustering within the mid-to-high registration range. This distribution suggests that the EREV market maintained a balance between consumer adoption and operational efficiency. The relatively wider range of energy intensity compared with that of BEVs may indicate a more uniform technological approach in EREV powertrain design or reflect a market orientation in which attributes such as performance and utility outweigh energy efficiency in shaping consumer preferences. For the PHEVs shown in Fig. 1 c, PHEV models with high registrations, such as BYD Destroyer 05 and BYD Song Pro, were concentrated in the lower-to-moderate energy intensity range. The large number of models represented in the PHEV category reflected a highly competitive segment. The observed gradient in energy intensity, combined with variations in registration volumes, suggests that automakers were actively competing to deliver PHEV models that balance electric mode efficiency with extended range capabilities via internal combustion engine support. The most successful models in terms of adoption were those that effectively achieved this trade-off.

[2] A real-world data validation of the proposed bottom-up model is provided in Appendix L.

Fig. 1 d aggregates data across all vehicle types to offer a macrolevel perspective. The analysis confirmed that mainstream models—such as the BYD Qin PLUS and Tesla Model Y—led in both registrations and favorable energy intensity, serving as market anchors. Their dominance reflected a combination of brand recognition, advanced energy management technologies, and competitive pricing strategies. In contrast, niche models such as the Tank 700 New Energy exhibited lower adoption and more varied energy consumption, suggesting a focus on specialized segments such as off-road or luxury markets, where energy efficiency may be secondary to other performance attributes.

Collectively, the above results provide valuable insights into consumer preferences, indicating that energy-efficient models tend to achieve greater market penetration. They also reflect the competitive dynamics of the EV market, where manufacturers had to balance technological advancements in energy management with market-facing factors such as brand positioning and pricing strategies. Moreover, the findings inform EV-related policymaking and marketing efforts by underscoring the pivotal role of energy efficiency in driving adoption across different EV segments.

Fig. 2 depicts electricity consumption trends across different types of EVs. Fig. 2 a shows the temporal dynamics of the electricity consumption of all EV operations, including the operations of BEVs, EREVs, and PHEVs, from June 2022 to December 2024 in Chongqing. The overall consumption exhibited an upward trajectory over time, albeit with noticeable fluctuations. The 95% confidence intervals for each vehicle type indicated varying degrees of volatility. From June 2022 to August 2023, BEVs consistently exhibited higher electricity consumption than EREVs and PHEVs (detailed results are provided in Appendix I). For example, during mid-2023, the electricity consumption curve for BEVs remained above those of EREVs and PHEVs. This pattern suggests that BEVs—which are fully reliant on battery-electric power—tended to exhibit higher energy use under normal operating conditions during the study period.

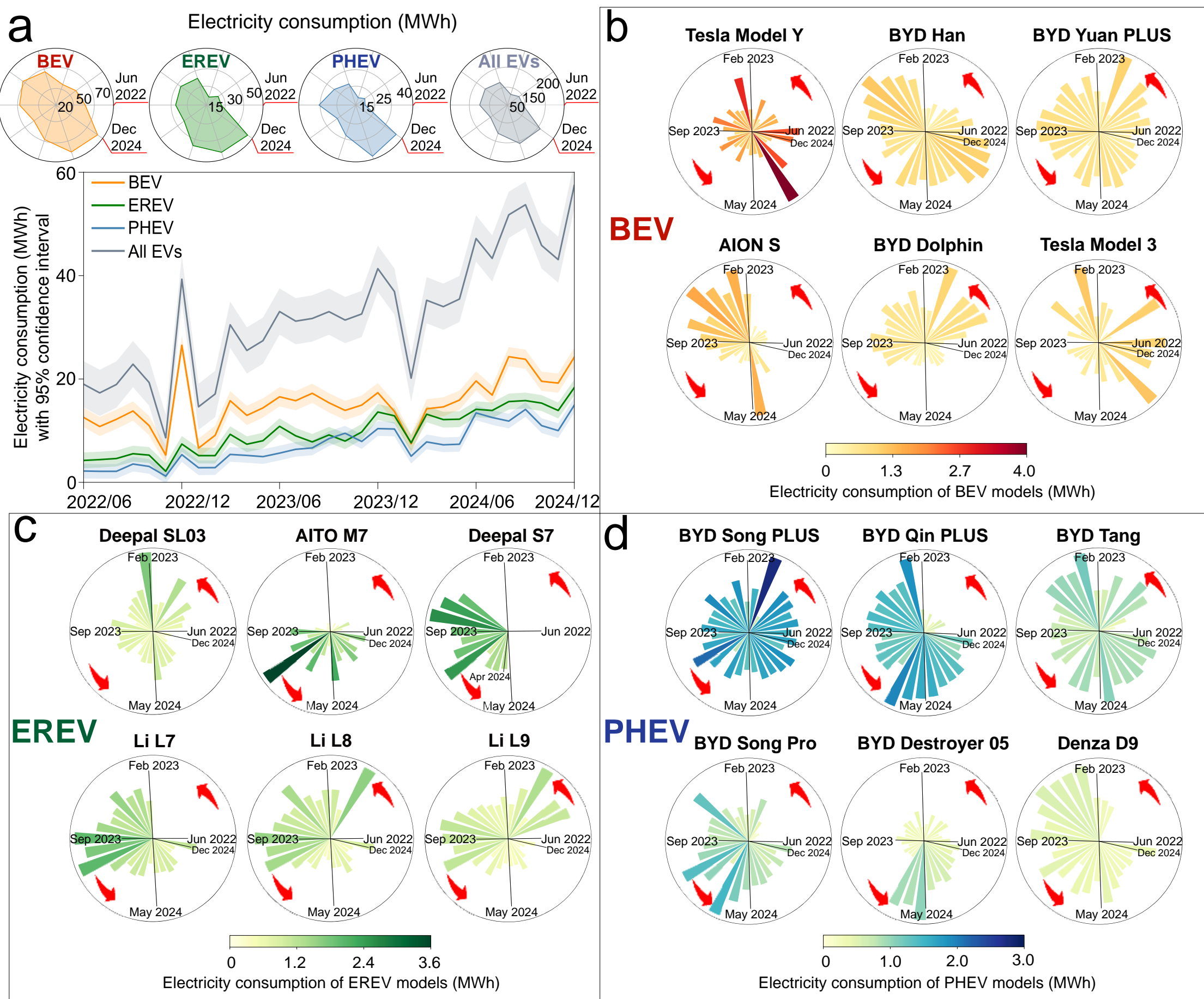

**Fig. 2.** Electricity consumption trends across different types of EVs from June 2022 to December 2024 in Chongqing: (a) temporal trends for BEVs, EREVs, PHEVs, and all EVs; (b–d) model-specific electricity consumption distributions for BEVs, EREVs, and PHEVs.

Fig. 2 b presents the model-specific electricity consumption distributions for six BEV models from June 2022 to December 2024 in Chongqing. Tesla Model Y displayed notable month-to-month variability, with relatively high consumption in certain months (e.g., May 2024). BYD Han and BYD Yuan PLUS also fluctuated across the study period. In contrast, Tesla Model 3 and AION S showed more moderate and stable consumption patterns. These variations in electricity consumption among BEV models may be attributed to differences in driving conditions (e.g., urban vs. highway use, temperature effects on battery efficiency), user behavior (e.g., travel distance and frequency), and underlying vehicle technologies such as battery systems and powertrain configurations.

For the EREV models (Fig. 2 c), with a consumption range of 0 to 3.6 megawatt-hour

(MWh), models such as Deepal SL03, AITO M7, Deepal S7, and the Li series (Li L7, L8, and L9) exhibited distinct consumption patterns. Specifically, AITO M7 demonstrated higher electricity use in specific months, whereas others, such as Li L7, showed more consistent consumption over time. The extended-range configuration of these vehicles may contribute to varying consumption behaviors. Factors such as the frequency of engine engagement, battery state-of-charge management strategies, and driving conditions could significantly influence electricity use. For example, in colder months, increased reliance on range-extending engines for cabins or battery heating may indirectly affect electricity consumption trends.

Fig. 2 d illustrates the electricity consumption of the top six PHEV models on a 0–3.0 MWh scale. Models from BYD—including Song PLUS, Qin PLUS, Tang, Song Pro, and Destroyer 05—along with Denza D9, exhibited clear consumption differences. The Tang registered higher peaks in certain months, whereas the Qin PLUS maintained a steadier level. As plug-in hybrids, these vehicles switch among electric-only, hybrid, and engine-only modes, so electricity use depends on battery capacity, powertrain efficiency, and driver mode selection. For example, a driver favoring electric-only commutes would show lower monthly consumption than one frequently using the hybrid mode for longer trips, a distinction reflected in the varied monthly patterns across PHEV models.

In summary, the above results provide valuable insights into the electricity consumption patterns of different EV types over the past three years in Chongqing, highlighting key opportunities for future research on optimizing energy management strategies—specifically, the capacity planning of EV charging infrastructure. These findings also provide a preliminary response to Question 1 outlined in Section 1.

### *3.2. Retrospective evaluation of operational carbon emissions for EV models*

This study retrospectively evaluated the operational carbon emissions of EV models over the past three years by combining bottom-up electricity consumption estimates with region-specific emission factors. Fig. 3 presents the trends in operational carbon emissions across different EV types in Chongqing from June 2022 to December 2024. Specifically, Fig. 3 a shows the overall trends in operational carbon emissions and the respective shares of BEVs, EREVs, and PHEVs from 2022–2024. Overall, emissions exhibited temporal fluctuations, reflecting the dynamic nature of carbon output contributions from different vehicle types over the study period. For example, EREVs demonstrated a noticeable increase in emissions around late 2022. This rise may have been driven by multiple factors, such as an uptick in long-distance travel, which—given EREVs' reliance on both electric and combustion power—would result in higher emissions. Alternatively, colder weather conditions during that period may have led to increased energy consumption for cabin heating, intensifying the operational burden and thereby increasing emissions.

The pie charts in Fig. 3 a reveal shifting proportional contributions: in 2022, BEVs accounted for 63.4% [76 kilotons of carbon dioxide (kt$CO_2$)], EREVs accounted for 23.1%, and PHEVs accounted for 13.5%. By 2023, total emissions increased to 183 kt$CO_2$, with BEVs contributing 48.7%, EREVs contributing 29.5%, and PHEVs contributing 21.8%. In 2024, total emissions reached 264 kt$CO_2$, with 42.3% for BEVs, 32.7% for EREVs, and 24.9% for PHEVs. These evolving shares suggest changes in vehicle utilization patterns—such as an increase in PHEV deployment for extended trips that activate their combustion mode or technological advancements that reduce BEV emissions per unit despite fleet expansion. Similarly, EREV emissions may have temporarily increased because of performance upgrades in combustion engines or energy-recovery systems, whereas PHEVs could have benefited from improved hybrid algorithms that enhance efficiency under specific conditions. Variations in traffic or driving environments where these vehicles were predominantly used may also have influenced these distributional shifts.

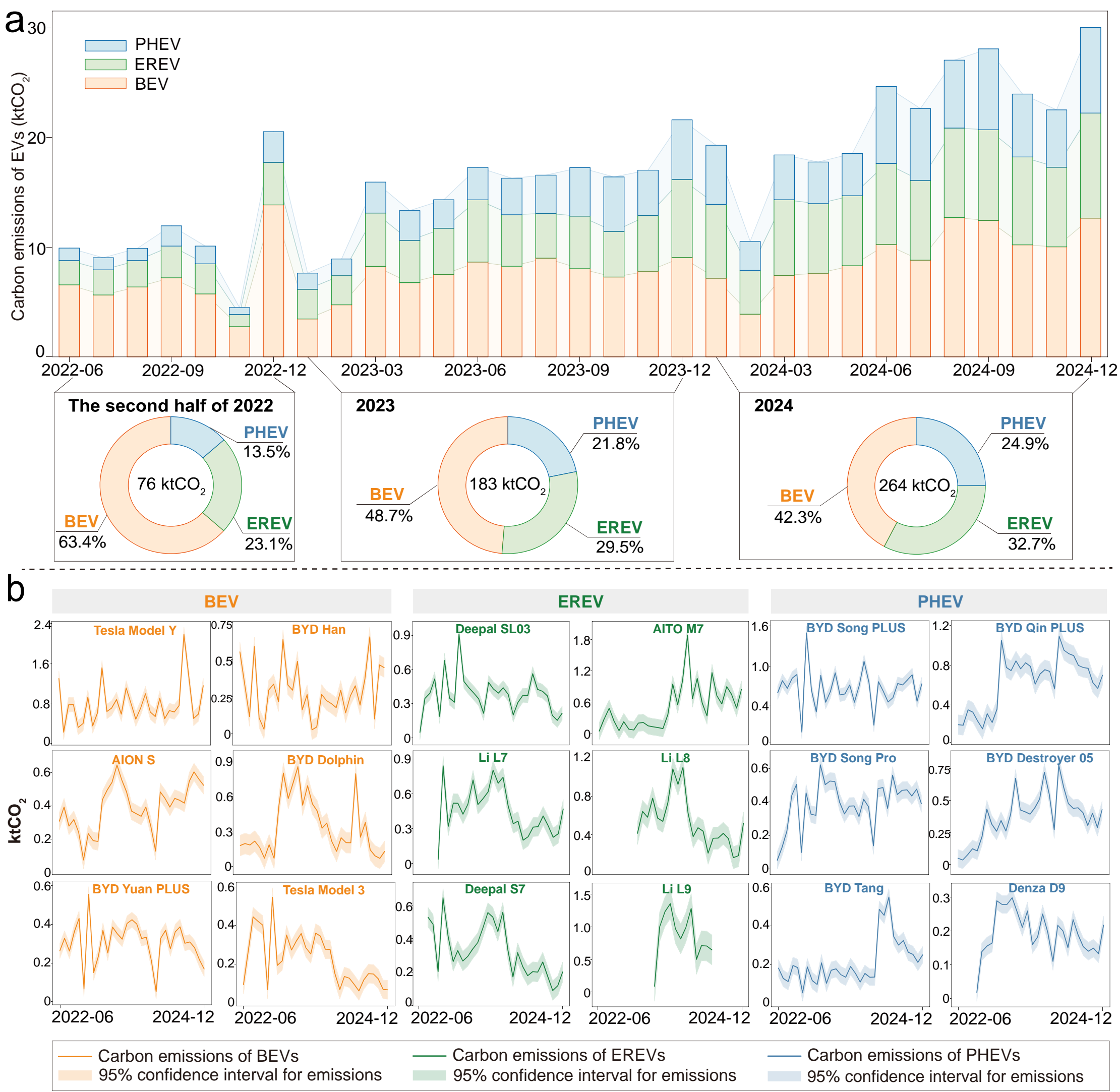

**Fig. 3.** Trends of operational carbon emissions among different EV types from June 2022 to December 2024 in Chongqing: (a) overall trends in operational carbon emissions and the emission shares of BEVs, EREVs, and PHEVs in the second half of 2022, 2023, and 2024; (b) monthly changes in operational carbon emissions for the top-selling EV models.

Fig. 3 b shows the monthly operational carbon emissions of the top six models for each EV type (BEV, EREV, PHEV) over the same period. Among BEVs, the Tesla Model Y exhibited higher and more volatile emissions than did models such as the BYD Dolphin, which presented lower and more stable patterns. These differences were attributed to design characteristics: the Tesla Model Y, often deployed for both urban and highway driving, likely featured a larger battery and higher-powered motors that increased energy consumption under load, thereby increasing upstream emissions from electricity generation. In contrast, the BYD Dolphin appeared more suited to urban commuting, potentially with a more compact design and a battery system optimized for short-range efficiency, resulting in consistently lower emissions. Variations in battery chemistry and management systems also influenced energy draw and, consequently, emissions. Furthermore, driver behavior played a role—Tesla Model Y users may have engaged in more dynamic driving styles, whereas BYD Dolphin users tended to adhere to calmer, efficiency-oriented usage.

For EREVs, models such as Deepal SL03 (with relatively fluctuating emissions) and AITO M7 (characterized by distinct variations in emission levels) demonstrated divergent trends shaped by powertrain configurations and energy management strategies. Deepal SL03 appeared to switch more frequently between pure electric and extended-range modes, resulting in greater variability in emissions. AITO M7, potentially equipped with a more complex energy management system designed to enhance performance under specific conditions, showed noticeable increases in emissions when the internal combustion engine was activated to provide supplementary power. These emission patterns are influenced by the design and efficiency of the internal combustion engine, as well as by the strategies governing transitions between energy sources and energy recuperation.

Among PHEVs, the BYD Song PLUS exhibited comparatively higher emissions during certain periods, whereas the BYD Destroyer 05 maintained more stable and moderate levels. The hybrid system efficiency served as a key differentiator. The BYD Song PLUS appeared to adopt a less optimized hybrid architecture, which may have led to less coordinated operation between the electric motor and the combustion engine, thereby increasing energy losses and emissions. Engine displacement also played a role; the larger engine of the Song PLUS consumed more fuel when engaged. Additionally, usage patterns contributed significantly—frequent operation in combustion engine-dominant modes (e.g., during long-distance travel without regular charging) increased emissions. In contrast, BYD Destroyer 05 users tended to rely more on electric-only operations for shorter trips or benefited from a better-balanced hybrid system, which helped maintain lower emissions.

Collectively, these results offer a comprehensive picture of the emission profiles for different EV types and leading models in Chongqing, providing a solid evidence base for policies to reduce the carbon footprint of EV operations. Detailed analysis of these profiles can guide the industry in adopting more targeted and effective measures, advancing a sustainable future for electric mobility in which both vehicle categories and individual models play a significant role in reducing transportation-related emissions. This work also lays the foundation for further research on EV charging demand—particularly for estimating the city-scale capacity requirements for charging infrastructure—and fully addresses Question 1 outlined in Section 1.

### *3.3. City-scale historical assessment of EV charging pile capacity*

Building on the findings in Sections 3.1 and 3.2, this section applies the optimization model described in Section 2.2 to estimate Chongqing's historical EV charging capacity, accounting for spatial heterogeneity across its districts and counties. Fig. 4 shows the spatial distribution of charging pile deployment across different regions of Chongqing. Fig. 4 a depicts the charging pile planning results for 38 districts and counties from 2022 to 2024. Over this period, a clear spatiotemporal evolution in infrastructure deployment was observed. In 2022, charging piles were relatively sparsely distributed, and only a few districts (e.g., Yubei, Jiulongpo, and Shapingba) exhibited moderately high volumes. By 2023 and 2024, the total number of charging piles increased markedly across the region. The core urban districts (e.g., Yubei, Jiangbei, and Nan'an), in particular, demonstrated more substantial growth. For example, Yubei experienced a substantial increase in charging piles in 2024 relative to earlier years, reflecting accelerated infrastructure development. This trend reflects a strategic and progressively intensified investment in charging infrastructure, likely driven by rising EV charging demand and supportive policy measures that promoted the transition to electric mobility.

Fig. 4 b presents the year-over-year growth of charging piles in nine key districts of Chongqing from 2022 to 2024. In Yubei District, the number of charging piles increased steadily, with an 83% growth from 2022 to 2023 and a 61% growth from 2023 to 2024, indicating rapid infrastructure expansion. Similarly, Jiangbei District experienced a 55% increase from 2022 to 2023 and a 60% increase from 2023 to 2024. Growth patterns varied across districts; for example, Beibei District recorded a 49% increase from 2022 to 2023 and a 45% increase from 2023 to 2024, reflecting a comparatively moderate increase. These differences in growth rates were likely influenced by factors such as the initial infrastructure base, local government investment priorities, and the underlying demand for EV charging (potentially shaped by population density, economic development, and the regional adoption of EVs).

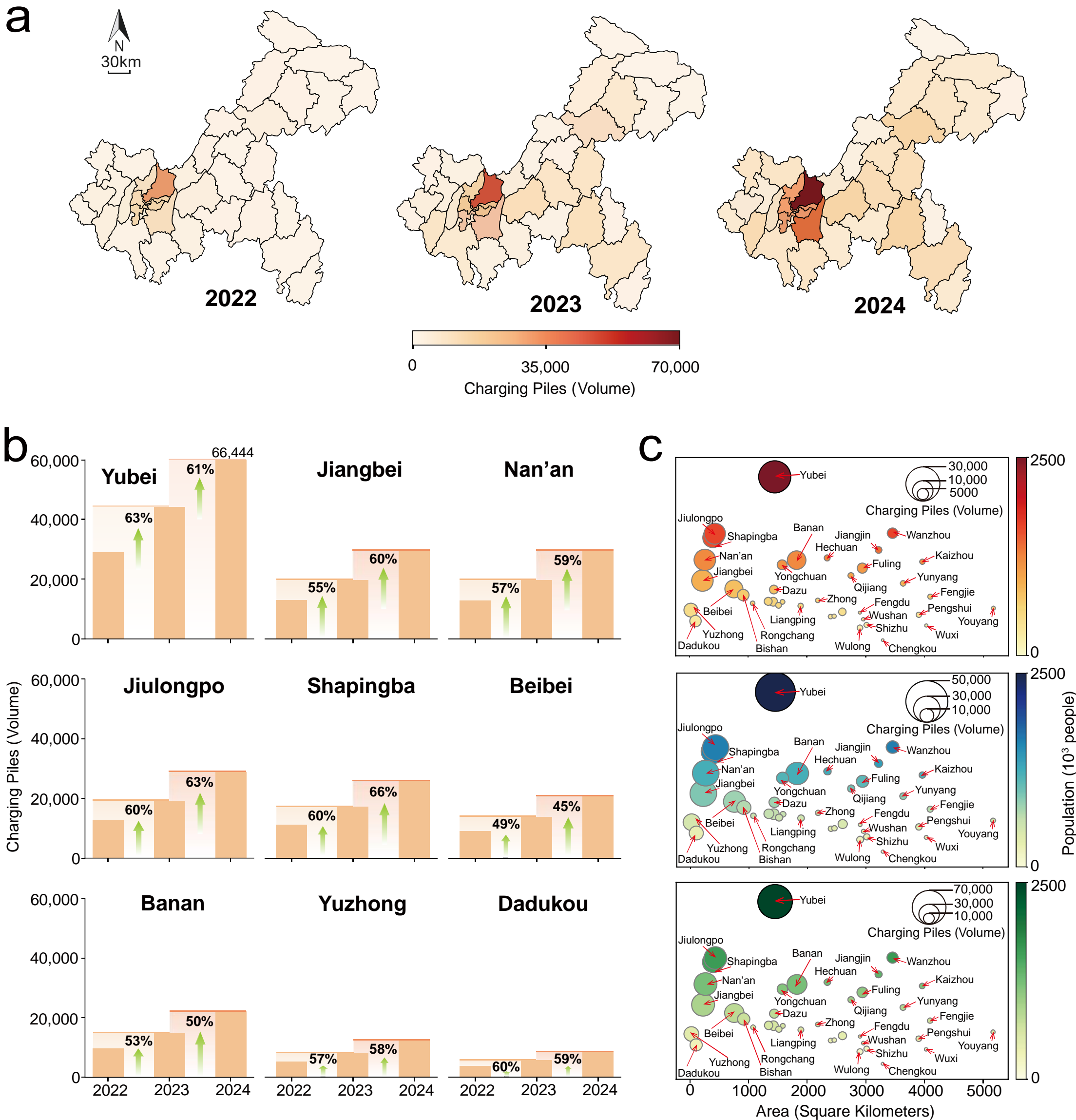

**Fig. 4.** Charging pile deployment across districts in Chongqing from 2022 to 2024: (a) spatiotemporal distribution of charging piles by district and county; (b) annual growth trends of charging piles by district and county; (c) correlation analysis between pile volume, population size, and district/county area.

Fig. 4 c illustrates the relationships among the pile volume, population size, and district/county area in Chongqing's districts from 2022 to 2024. The distribution in 2022 revealed that charging piles were unevenly distributed, with certain districts (e.g., Yubei, Jiulongpo, and Shapingba) exhibiting a relatively high concentration of charging infrastructure. The distribution in 2023 illustrated the population spread and showed a partial correlation between population density and the number of charging piles. Districts with higher population densities (e.g., Yubei) generally had more charging piles, although the relationship was not strictly linear. The distribution in 2024 demonstrated the geographic size of each district and indicated that districts with larger land areas (e.g., Youyang) did not necessarily have more charging piles. These observations suggest that the deployment of charging infrastructure in Chongqing was influenced by a combination of factors (including population demand, land use patterns, and strategic planning).

In summary, these findings collectively offer a comprehensive perspective on the deployment of charging piles across Chongqing's districts and counties from 2022 to 2024. The identified spatiotemporal patterns, growth trajectories, and multidimensional relationships provide valuable insights for optimizing future charging infrastructure planning, deepening the understanding of EV development dynamics. Overall, the above results provide a direct response to Question 2 raised in Section 1.

## 4. Discussion

This section discusses several critical aspects of the study, including short-term projections of EV charging demand at the city scale, as well as city-level planning for future EV charging pile capacity. Additionally, the research implications were addressed.

### *4.1. City-scale projected charging demand and charging pile capacity planning*

This study projected future EV electricity demand and planned charging infrastructure capacity, with a particular focus on city-level charging pile capacity. Fig. 5 shows the projected electricity demand trends for different EV types in Chongqing from 2025 to 2030. Specifically, Fig. 5 a presents the quarterly electricity demand trends for the operation of BEVs, EREVs, and PHEVs during this period: BEVs would reveal a clear trend of gradually increasing consumption across successive quarters. In earlier periods, such as March–May 2025, consumption values were relatively low, but by March–May 2028, a notable upward shift would be observed, with later quarters consistently exhibiting greater electricity use. This pattern suggests that BEV adoption and usage intensity have increased over time, resulting in elevated energy demand. EREVs would also show quarterly fluctuations; however, the overall consumption levels would remain lower than those of BEVs. Like EREVs, PHEVs would exhibit quarterly variability; however, their electricity consumption would remain lower and more constrained than that of BEVs does, reflecting the moderate demand profile characteristic of plug-in hybrid configurations.

Fig. 5 b shows the distributions of total electricity demand for BEVs, EREVs, and PHEVs in Chongqing. BEVs would exhibit the widest distribution, as indicated by longer whiskers, larger interquartile ranges, and a higher median consumption of approximately 100–200 gigawatt-hours (GWh). This suggests substantial variability in BEV usage—some vehicles are deployed for long-distance, high-energy trips, whereas others are used primarily for shorter commutes. In contrast, EREVs would show a more concentrated distribution, with a median of approximately 50–100 GWh, indicating mixed usage patterns: from short, electric-only trips with lower consumption to longer, hybrid-mode journeys with higher but less extreme consumption than BEVs. PHEVs would fall between BEVs and EREVs, reflecting their transitional role in the powertrain shift from EREVs to BEVs.

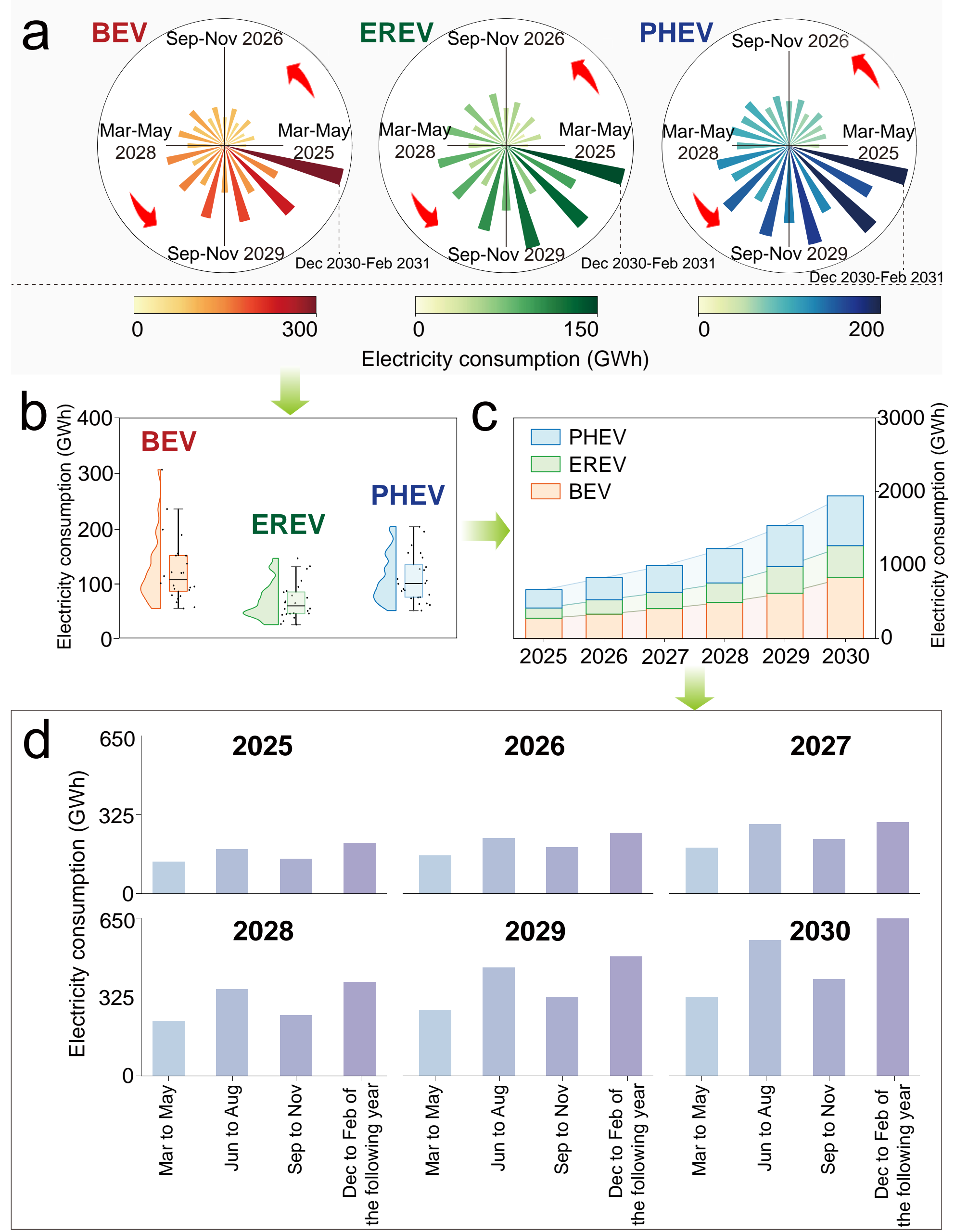

**Fig. 5.** Electricity demand trends across different types of EVs from 2025 to 2030 in Chongqing: (a) quarterly trends for BEVs, EREVs, and PHEVs; (b) distributions of electricity demand for BEVs, EREVs, and PHEVs; (c) annual growth in electricity demand for BEVs, EREVs, and PHEVs; (d) monthly changes in electricity demand for EV operations.

Fig. 5 c illustrates the annual electricity consumption trends for BEVs, EREVs, and PHEVs from 2025 to 2030 in Chongqing, revealing a consistent upward trajectory across all vehicle types. BEVs would account for the largest share of total consumption, with their contribution increasing steadily over time. This growing dominance results from factors such as rising BEV market penetration, advancements in battery technology enabling extended driving ranges, and a shift in consumer preferences toward fully EVs. Although electricity consumption by EREVs and PHEVs would also increase, their growth rates would remain relatively moderate. Overall, the collective upward trend highlights a surge in demand for EV-related electricity, driven by policy incentives promoting electric mobility, improvements in charging infrastructure, and heightened environmental awareness among consumers—factors that would accelerate the adoption of all electrified vehicle types.

Fig. 5 d presents the monthly electricity consumption distributions from 2025 to 2030 in Chongqing, grouped by meteorological season: spring (March–May), summer (June–August), autumn (September–November), and winter (December–February). Consumption patterns would clearly vary across seasons and years. In later years (particularly 2030), elevated consumption would be observed during summer and autumn for most vehicle types. These seasonal fluctuations indicate the potential impact of travel behavior and environmental conditions on electricity demand. For example, summer is typically associated with increased long-distance travel due to holidays, which would lead to more frequent and extended trips, thereby increasing energy use. Seasonal temperature extremes would also influence battery performance and necessitate greater energy use for in-vehicle climate control. Furthermore, electricity pricing policies aimed at demand-side management—such as seasonal rate incentives—would encourage increased charging activity during specific periods, contributing to the observed surges in consumption.

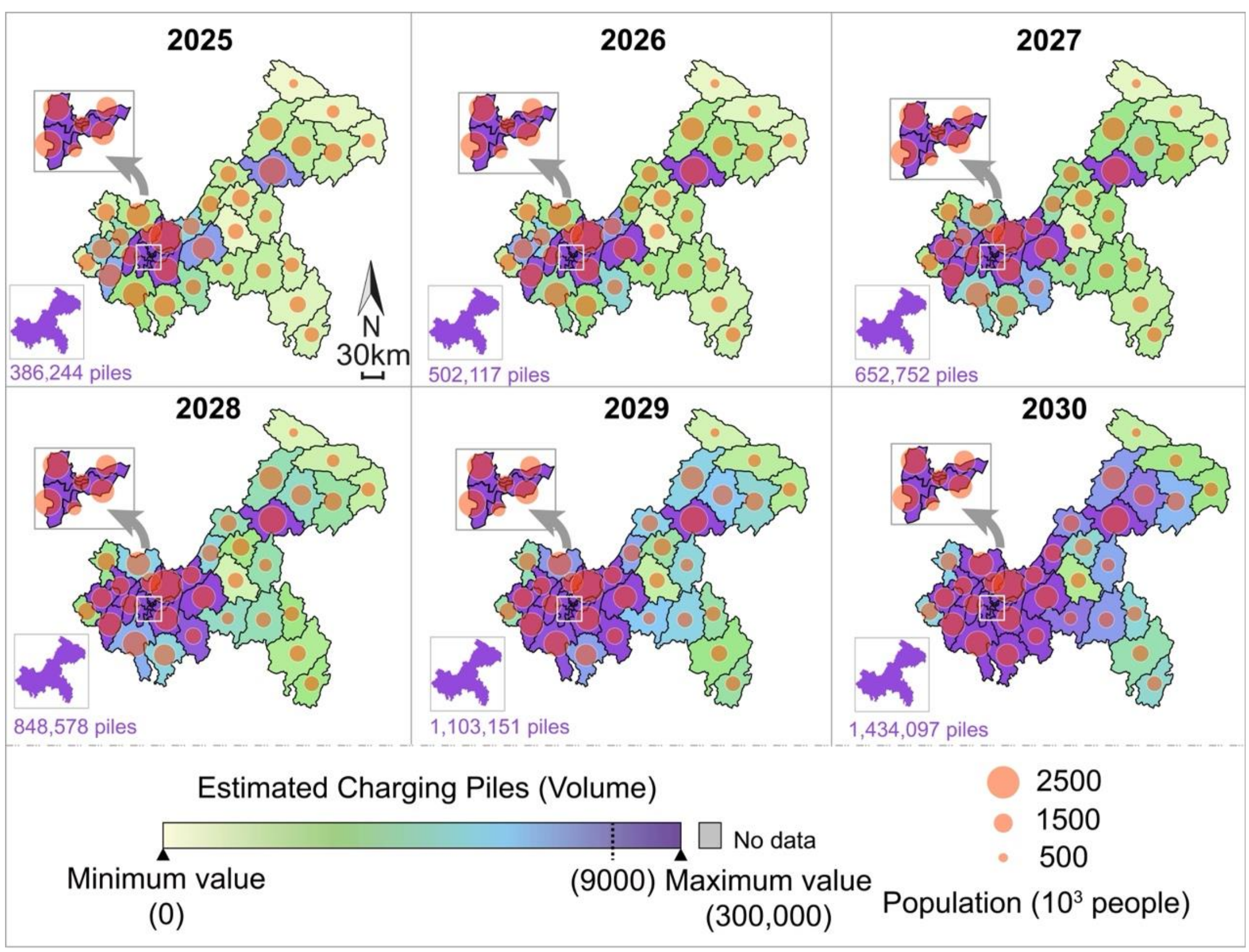

**Fig. 6.** Charging pile deployment planning in districts and counties of Chongqing: spatiotemporal distribution of piles and population development from 2025 to 2030.

Fig. 6 shows the projected deployment plan for charging piles across 38 districts and counties in Chongqing from 2025 to 2030, highlighting the spatiotemporal distribution of charging infrastructure and its correlation with population density (A sensitivity analysis examining the impact of EV registration changes on the future spatiotemporal distribution of charging piles in Chongqing is provided in Appendix J). From 2025 to 2030, charging piles would exhibit a strong and sustained upward trajectory citywide. In 2025, the total number of charging piles would reach 386,244, with a relatively sparse distribution [only a few core districts (e.g., Yubei, Jiulongpo, and Nan'an) would show moderate densities]. By 2026, this number would rise to 502,117, reflecting an initial expansion phase likely driven by policy incentives and growing market demand for EV infrastructure. This rapid growth would signal a strategic commitment to EV infrastructure development, in line with global decarbonization efforts and local transportation electrification goals. The compound annual growth rate derived from these figures would underscore the scale of investment required to support mass EV adoption in the region.

Spatially, the growth of charging piles would be uneven and closely follow population density patterns. Core urban districts (e.g., Yubei, Jiangbei, Nan'an, Jiulongpo, and Shapingba) would receive a disproportionate share of infrastructure investment. For example, densely populated central zones (e.g., Yubei, Jiulongpo, and Shapingba) would consistently show higher allocations, approaching the upper limit of 30,000 charging piles in later years such as 2030. This pattern would demonstrate a demand-driven planning approach, whereby areas with higher population concentrations—and consequently greater potential EV ownership—would be prioritized for infrastructure deployment.

In 2025, only a few districts (e.g., Wanzhou, Yongchuan, Fuling, and Changshou) would approach moderate density levels (e.g., 5000–10,000 piles). By 2030, the vast majority of districts (e.g., Wanzhou, Bishan, Fuling, and Dadukou) across Chongqing would exhibit high charging pile volumes, with many nearing or reaching the 30,000-pile mark. Such widespread densification would reflect a mature infrastructure network capable of supporting a large-scale shift to electric mobility. Moreover, spatial expansion would reveal more nuanced patterns: peripheral districts (e.g., Banan, Beibei, and Qijiang) would gradually acquire charging infrastructure over time, suggesting a phased rollout strategy that prioritized high-demand urban centers initially, followed by expansion into suburban and rural areas to ensure equitable access to EV services. This approach balances the immediate need to serve densely populated regions with the long-term objective of establishing a comprehensive and inclusive regional charging network.[3]

In summary, projections for Chongqing from 2025 to 2030 indicate steadily rising electricity demand across BEVs, EREVs, and PHEVs, with BEVs contributing the most and most variable consumption due to increasing adoption, technological advancements, and shifting consumer preferences. Seasonal and quarterly variations reflect the impacts of travel behavior, climate conditions, and demand-side management on electricity use. Moreover, the planned deployment of charging piles shows a strong citywide growth trend, with expansion closely aligned with population density. Core urban districts (e.g., Yubei) would initially receive the bulk of infrastructure investment, followed by phased rollout to peripheral areas, illustrating a demand-driven strategy to support large-scale EV adoption. Collectively, these findings offer valuable insights for optimizing future charging infrastructure planning and managing projected electricity demand, directly addressing Question 3 raised in Section 1.

[3] An integrated capacity planning and spatial site-selection analysis is provided in Appendix M.

### *4.2. Development of a city-scale EV charging model with global applicability*

Transitioning toward an electrified transportation system requires not only technological innovation but also proactive, adaptive infrastructure planning [51, 52]. While Chongqing was selected as the case study for analyzing EV charging demand and infrastructure capacity planning, the proposed two-stage framework—coupling demand-side estimation with optimization-based capacity planning (Fig. 7)—is designed to be transferable to other urban contexts worldwide. In other words, the approach is adaptable for application across cities with diverse socio-economic, geographic, and policy conditions, enabling meaningful cross-regional comparisons. This framework provides policymakers with a versatile decision-support tool to facilitate sustainable and cost-effective EV infrastructure deployment aligned with low-carbon transportation electrification goals.

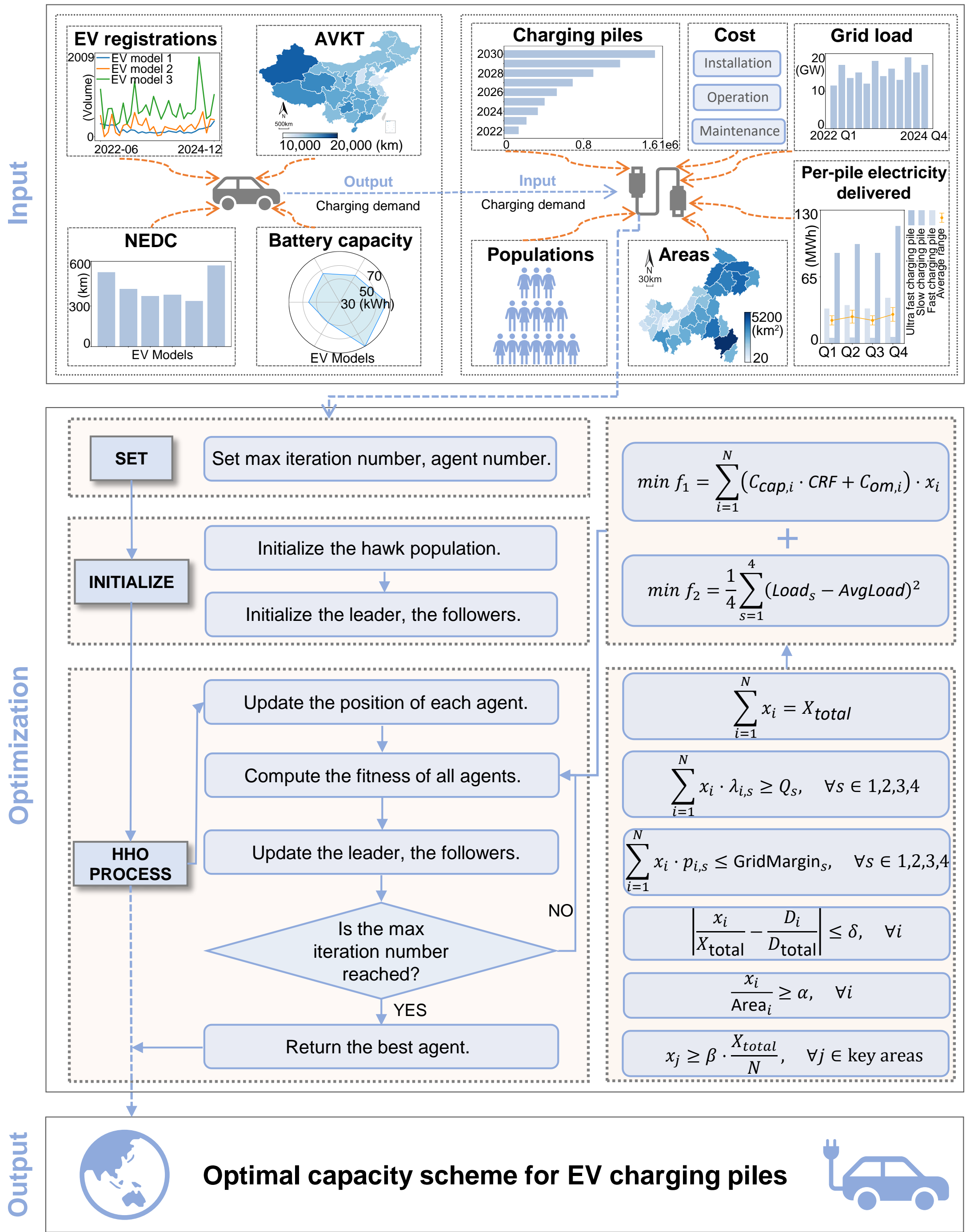

**Fig. 7.** Operational workflow of the city-scale capacity planning model for EV charging piles.

As illustrated in Fig. 7, the workflow of the proposed model integrated diverse inputs, such as EV registrations, EV battery capacity, charging piles, populations, and grid loads. The framework first estimated charging demand via a bottom-up model and subsequently integrated the estimated demand into a multi-objective optimization model. Solved via the HHO algorithm, this model simultaneously minimized total costs and load fluctuations, producing spatially resolved, seasonally adaptive siting and sizing strategies for charging piles. A post-processing feasibility screening is also provided in Appendix K. To ensure its applicability beyond Chongqing, the framework was designed to accommodate city-specific inputs and planning constraints. For example, cities with high adoption rates of private EVs may place more emphasis on residential slow-charging deployment, whereas megacities facing limited land availability may prioritize ultrafast charging at transport hubs.

International experience reinforces the need for integrated, forward-looking planning [53]. In the Netherlands and Norway, charging infrastructure deployment is closely coordinated with renewable energy integration and parking policies [54]. In contrast, cities in the United States, such as Los Angeles and New York, focus on addressing equity and accessibility gaps in disadvantaged communities [55]. In emerging economies such as India and Brazil, public–private partnerships and fiscal incentives play a central role in infrastructure scaling [56]. In addition to modeling and optimization, effective city-scale planning should be complemented by policy instruments such as the following:

- **National-level policy alignment and standardization.** Establishing unified national frameworks for EV charging infrastructure is essential for ensuring interoperability and coordinated development across regions [57, 58]. This includes standardized guidelines for charging pile construction, operation, and safety [59]. For example, the European Union has made significant efforts toward harmonizing charging protocols across member states, enabling cross-border EV mobility and fostering a more integrated EV market [60]. Such frameworks not only reduce institutional fragmentation but also facilitate the large-scale application of planning tools such as the proposed model by providing consistent technical and regulatory baselines [61, 62].
- **Targeted incentives for both private and public stakeholders.** Effective scaling of charging infrastructure requires the mobilization of both private capital and public investment [63]. Incentive mechanisms (such as tax credits [64], installation subsidies [65], and preferential financing [66]) can reduce upfront costs for private actors. In the United States, for example, several state governments offer financial incentives to businesses that deploy EV charging stations [67]. On the public side, targeted investment in underserved areas (e.g., rural and peri-urban zones) can address spatial inequities and catalyze demand. These demand-side effects, in turn, enhance the relevance and necessity of data-driven capacity planning approaches.

- **Integration with smart grids and renewable energy systems.** Future-proof planning requires close alignment between the EV infrastructure and the evolving energy ecosystem [68]. Smart grid technologies offer dynamic load management and real-time demand response capabilities, which can be directly incorporated into the optimization module of planning tools. Countries such as Denmark are exploring integrated charging strategies that coordinate with wind and solar energy availability [69]. Moreover, policy support for the vehicle-to-grid technologies—whereby EVs act as distributed energy storage—can further augment grid flexibility [70, 71]. Accounting for such bidirectional energy flows within capacity planning models ensures more resilient and adaptive infrastructure deployment.

Collectively, these policy instruments not only create an enabling environment for infrastructure deployment but also enhance the practical utility of integrated planning models. By aligning data-driven optimization frameworks with real-world regulatory, financial, and technological contexts, cities can achieve more balanced, cost-effective, and future-resilient EV charging networks. As urban mobility systems continue to electrify, the integration of planning tools with multilevel policy design will be critical for translating long-term decarbonization goals into actionable infrastructure strategies.

## 5. Conclusion

This study presented a generalizable modeling framework to analyze historical and future EV charging demand and to plan corresponding infrastructure capacity at the city scale. The framework adopted a bottom-up approach to estimate charging demand and associated carbon emissions, integrating diverse data sources spanning socio-economic factors, technological attributes, and climatic conditions. Subsequently, the estimated electricity demand was integrated into a multi-objective model to optimize the charging pile capacity by minimizing costs and balancing the grid load, with solutions generated via the HHO algorithm. A case city in Chongqing, was investigated, as it stands out as a rapidly growing EV industry cluster in China, featuring a strong industrial base, supportive policies, and diverse urban morphologies. The analytical period in Chongqing spanned from 2022 to 2030 and covered both central urban districts and peripheral counties, offering a comprehensive testbed for validating the framework. The key findings of this study are listed below.

### *5.1. Key findings*

- **The electricity consumption of Chongqing's EV operations showed pronounced monthly fluctuations, increasing from 18.9 GWh in June 2022 to 57.5 GWh in December 2024 (a 204% rise), with corresponding carbon emissions increasing from 9.9 kt$CO_2$ to 30.0 kt$CO_2$.** The monthly consumption peak occurred in December 2024 (57.5 GWh), driven by winter-related demand, such as cabin heating and battery heat-pump operation. BEVs remained the largest contributor, with monthly consumption increasing from 12.6 GWh to 24.2 GWh (a 92.0% increase); however, their share decreased from 66.7% to 42.1% due to faster growth in other powertrain types. EREV and PHEV consumption increased from 4.2 GWh to 18.3 GWh and from 2.2 GWh to 14.9 GWh, respectively—growth rates exceeding 350%. From June 2023 onward, the combined monthly consumption of EREVs and PHEVs began to rival that of BEVs, indicating increasing technological diversity. While BEVs still constituted the largest share of emissions, EREVs and PHEVs together accounted for more than 57% of EV-related monthly emissions in the second half of 2024, underscoring the need for powertrain-specific decarbonization strategies.

- **From 2022 to 2024, Chongqing planned an additional 181,622 charging piles—an average annual growth rate of 38%—with deployment concentrated in dense, high-demand urban districts.** Yubei, Jiangbei, Nan'an, and Jiulongpo consistently ranked among the top in terms of cumulative capacity, together accounting for more than one-third of the city's total installations by 2024. Yubei alone hosted 66,444 charging piles—over 64 times more than Youyang County did, which received only 1,028 charging piles despite covering an area more than 3.6 times larger. This disparity underscores that infrastructure planning was guided primarily by demand intensity rather than geographic coverage. Densely populated districts received disproportionately higher capacity allocations despite their limited spatial extent, whereas geographically expansive but sparsely populated counties experienced slower growth and lower installation levels. This pattern reflects a demand-driven planning approach and highlights the need for adaptive strategies in peripheral areas to mitigate emerging service disparities.
- **From 2025 to 2030, Chongqing's EV electricity demand was projected to rise from 668 to 1940 GWh, averaging 23.8% annual growth, while charging infrastructure was expected to exceed 1.4 million piles by 2030.** The highest seasonal demand was predicted for winter 2030, with BEV consumption reaching 306.2 GWh—up from 54.5 GWh in spring 2025, an 82% increase. Although BEVs would remain the largest single category, the combined demand from EREVs and PHEVs would be anticipated to surpass that of BEVs during 2025–2030, indicating a more diversified EV fleet. The charging capacity would remain concentrated in high-demand urban districts such as Yubei, Jiangbei, Nan'an, and Jiulongpo, with Yubei alone projected to host over 296,042 piles by 2030. In contrast, remote counties were expected to experience slower growth, underscoring the need for targeted investment and planning to achieve a more balanced infrastructure distribution across the city.

### 5.2. Upcoming work

Building on the present analysis of EV charging demand and infrastructure capacity planning in Chongqing, future research will expand to a global scale. Although Chongqing serves as a representative case of a rapidly growing EV industry cluster in emerging economies, the developed modeling framework can be applied to multiple cities worldwide to enable cross-regional comparisons under diverse socio-economic, geographic, and policy conditions. To strengthen its adaptability across broader contexts, the next phase will advance in three main directions. First, the optimization model will be refined to incorporate carbon emissions minimization as a core objective alongside cost reduction and load balancing, enabling the assessment of decarbonization strategies aligned with varying national energy mixes and climate commitments. Second, scenario-based simulations will be conducted to evaluate the effects of different EV adoption patterns and policy interventions across cities. By capturing heterogeneity in charging demand and local socio-economic conditions, the study aims to identify best practices that are globally relevant yet adaptable to specific local contexts. Finally, a key future direction involves the explicit integration of human travel behavior, particularly inter-zonal trips, into the charging demand model. Building a unified framework that combines dynamic mobility patterns with static spatial factors will provide a more granular and behaviorally realistic foundation for infrastructure siting.

## Appendix

Please find the appendix in the supplementary materials (e-component).

## Acknowledgments

The first author appreciates the National Planning Office for Philosophy and Social Science Foundation of China (24BJY129) and Chongqing Association for Science and Technology Think Tank Research Project (2025KXKT40). The coauthors from Lawrence Berkeley National Laboratory declare that this manuscript was authored by an author at Lawrence Berkeley National Laboratory under Contract No. DE-AC02-05CH11231 with the U.S. Department of Energy. The U.S. Government retains, and the publisher, by accepting the article for publication, acknowledges, that the U.S. Government retains a non-exclusive, paid-up, irrevocable, world-wide license to publish or reproduce the published form of this manuscript, or allows others to do so, for U.S. Government purposes.

## Declaration of interests

The authors declare that they have no competing interests.